\documentstyle[preprint,aps]{revtex}

\input epsf
\epsfverbosetrue
\begin{document}

\draft
\preprint{UBCTP-95-05~~~ hep-th/yymmdd}
\title{Mean Field Description of the Fractional Quantum Hall Effect \\ Near 
$\nu=1/(2k+1)$} 

\vskip .5in

\author{Stephanie Curnoe$^1$ and Nathan Weiss$^2$}

\vskip .2in

\address{$^{1,2}$Department of Physics, University of British Columbia,\\
Vancouver, B.C. V6T 1Z1, Canada\\ and \\ $^2$Department of Particle Physics\\
Weizmann Institute of Science\\ Rehovot, Israel}
\maketitle
\vskip .3in

\begin{abstract}
The nature of Mean Field Solutions to the Equations of Motion of 
the Chern--Simons Landau--Ginsberg (CSLG) description of the
Fractional Quantum Hall Effect (FQHE) is studied. Beginning with
the conventional description of this model at some chemical
potential $\mu_0$ and magnetic field $B$ corresponding to
a ``special'' filling fraction $\nu=2\pi\rho/eB=1/n$ ($n=1,3,5\cdot
\cdot\cdot$) we show that a deviation of $\mu$ in a finite
range around $\mu_0$ 
does not change the Mean Field solution and thus the mean density
of particles in the model. This result holds not only for the
lowest energy Mean Field solution but for the vortex excitations
as well. The vortex configurations do not depend on $\mu$ in a finite
range about $\mu_0$ in
this model. However when $\mu-\mu_0 < \mu_{cr}^-$ (or 
$\mu-\mu_0>\mu_{cr}^+$) the lowest energy Mean Field solution
describes a condensate of vortices (or antivortices). We 
give numerical examples of vortex and antivortex configurations
and discuss the range of $\mu$ and $\nu$ over which the system
of vortices is dilute.
\end{abstract}


\vskip .5in

\narrowtext

\label{sec:intro}

There have been many interesting approaches to understanding the Fractional
Quantum Hall Effect (FQHE) in which a two--dimensional electron gas is 
subjected to a large transverse magnetic field.  One such approach which
is often called the ``Chern-Simons-Landau-Ginzberg'' (CSLG) approach to 
the FQHE \cite{kn:girvin1}-\cite{kn:jain1} makes use of the fact that
in two--dimensions a fermion (such as the electron) can be treated
as a boson to which is attached an odd number of magnetic flux
quanta. This magnetic flux can be the usual magnetic flux but more
conventionally one introduces an additional Gauge field known as
the ``statistical'' gauge field. The electron is then viewed as
a boson with an integer number of flux quanta of this ``statistical
magnetic field''. This  ``binding'' of the statistical flux to the
boson is achieved, technically, by the Chern--Simons interaction
which is briefly reviewed below. The key observation in this
approach to the FQHE is that the
most evident plateaus occur when the (conventional) magnetic
flux per particle is an odd integer:
\equation
	\nu=\frac{2\pi\rho}{eB}=\frac{1}{n}~~~~~~~~~
	~n=1,3,5\cdot\cdot\cdot
	\label{eq:a1}
\endequation
Thus if we describe each electron as a boson to which is attached
an odd number of statistical flux quanta, then the mean statistical
magnetic field will cancel the applied magnetic field precisely when
Eq. \ref{eq:a1} is satisfied. The resulting theory is that of
an interacting system of bosons with no net applied field which will
Bose condense. The resulting phenomenology has been studied in detail
in Refs. \cite{kn:girvin1}-\cite{kn:jain1}.

There has also been considerable work on describing the Fractional
Quantum Hall system when the density (or the magnetic field) is perturbed
away from the special filling fraction described by Eq. \ref{eq:a1}.
In the CSLG description
vortices occur in the system to  accommodate these fluctuations in the
density or in the magnetic field. It has been shown that
such vortices {\bf do} exist and they can be found by  perturbing the 
ground state of the system  at the special filling fraction by
adding (or removing) particles to the system.  In this way single
vortex solutions  can be found \cite{kn:girvin1},\cite{kn:zhang1}. 
These vortices are in fact the Laughlin quasi-particles and
quasi-holes \cite{kn:laugh1}.  When they occur in pairs they are neutral
excitations analogous to the 
rotons of 3-D superfluids. 

Vortices also form when the external magnetic 
field is changed
with the number of particles kept fixed.  In this case the 
perturbation is
a constant, uniform shift of the  field rather than a local density perturbation. 
Thus, in the limit of infinite volume,
a lattice of vortices covering the entire space is expected to 
form just as
would happen if the mean
particle density would be perturbed from that of Eq. (\ref{eq:a1}).
The conventional way to analyze this system is to begin at the special
filling fraction (i.e. to consider first the case in which the density
has been modified so that $2\pi\rho/eB=1/n$) and to accommodate the excess
density by means of these vortex excitations.
In this paper we shall take a slightly different approach. The electron 
density is described by means of a chemical potential $\mu$. When the
magnetic field is changed away from the special filling fraction 
we shall try to find the Mean Field ground state of the resulting 
system by looking for a solution to the resulting equations of motion. 
Since, in this case,
the mean Statistical Field
does not cancel the applied field there will be no uniform mean 
field solution
and the Mean Field equations are extremely ``frustrated''. We shall
discuss the properties of solutions to this system of equations 
analytically
and then present some numerical solutions. 

Our main goal in this paper is to display some novel features of the 
CSLG description
of the FQH system. To this end we adopt the simplified version of the
model which is described in Ref. \cite{kn:zhang2} which allows us to
show these features more clearly. We begin with a gas of bosons
represented by a (nonrelativistic) scalar field $\psi$ with a mass 
$m$. The fact
that these bosons are at some finite density $\rho$ will be 
implemented by considering
the system at some chemical potential $\mu$. The system will 
also have a Statistical
Gauge Field $a_\mu$ whose main purpose is to attach an integer 
number of statistical
magnetic flux quanta to each boson thus allowing them to 
describe an 
electron gas.
Our main simplification relative to the actual physical 
situation is to imagine
that these bosons have a hard core repulsive interaction 
($\propto \vert\psi\vert^4$)
instead of a Coulomb interaction. These bosons are also subjected 
to a fixed external
magnetic field $B$ which is described by an electromagnetic 
potential $A_\mu$.
The Lagrangian for this system is given (in units for which 
$\hbar=c=1$) by:
\begin{eqnarray}
    {\cal L} & = & -i\psi^{\dagger}(\partial_{0}-i(eA_{0}-ga_{0})\psi
	       + \frac{1}{2m}\vert(\nabla-i(e{\bf A}-{g\bf a}))\psi\vert^2 
	\nonumber \\
             & & - \frac{g^2}{4\pi n}\varepsilon_{\mu\nu\lambda}a^{\mu}
		 \partial^{\nu}a^{\lambda}
               + \frac{\lambda}{2}(\psi^{\dagger}\psi)^2
               -\mu\psi^{\dagger}\psi  
	\label{eq:lagrange}
\end{eqnarray}
We shall see in a moment that in order for $\psi$ to describe electrons
$n$ must be an odd integer.
The third term in this Lagrangian
is the Chern-Simons term which is responsible for the fact that Statistical
Magnetic Flux is attached to each particle. The equation of motion for the 
field $a_0$,
\equation
	\frac{g}{2\pi n} b~=~ \psi^\dagger\psi
\endequation
implies that to each particle (with $\int d^2x \psi^\dagger\psi =1$) is attached
a Statistical magnetic flux $g\Phi=2\pi n$.  Thus if $n$ is an odd integer
each boson has an odd number of Statistical flux quanta attached to 
it so that
it has Fermi statistics \cite{kn:transmut}.

In the special case when the filling fraction
\equation
	\nu~=~\frac{ 2\pi\rho}{eB}~=~\frac{1}{n}
\endequation
(where $\rho=\psi^\dagger\psi$ is the density of particles)
the  average effect of the Statistical field precisely cancels that 
of the
externally applied magnetic field i.e. $g<b>=eB$ so that a gauge can 
be chosen
in which, on the average, $g<a_i>=eA_i$.  This, of course, corresponds to 
the special filling fraction and leads to the CSLG description of the Fractional
Quantum Hall Effect.

We begin our discussion by writing down the equations of motion which
result from varying $\psi$, $a_0$ and $a_i$ in the Lagrangian (\ref{eq:lagrange}).
Since $A_\mu$ is the externally applied Gauge Potential due to a 
spatially
constant magnetic field we choose $A_0=0$.
\equation
	i(\partial_{0} +iga_{0})\psi + \frac{D_i^2}{2m}\psi 	
	-\lambda(\psi^{\dagger}\psi)\psi + \mu\psi = 0  
	\label{eq:motion1}
\endequation
\equation
	\psi^{\dagger}\psi =  \frac{gb}{2\pi n} \label{eq:motion2}
\endequation
\equation
	\frac{i}{2m}{(\psi^{\dagger}D_{i}\psi - (D_{i}\psi)^{\dagger}\psi)}
	=\frac{g}{2\pi n}
	{\varepsilon_{ij}(-\partial_{0}a_{j}+ \partial_{j}a_{0})}
	\equiv \frac{g}{2\pi n}{\varepsilon_{ij}e^{j}}
	\label{eq:motion3}
\endequation
where ${ D_i} = \partial_i - ie{A_i}+ig{a_i}$ is the covariant derivative
and $e_j$ is the statistical electric field.
The first equation (\ref{eq:motion1}) is the generalized (nonlinear) 
Shroedinger
equation for this system. The second
equation (\ref{eq:motion2}) implements the constraint that there are $n$ 
flux quanta per particle as discussed above. The third equation which also results from the
Chern--Simons term in the action relates the statistical electric field
to the particle current.

The above equations of motion are {\bf operator} equations for the
Quantum Fields $\psi$ and $a_i$. The technique  which we 
use to study these Quantum Equations (see Ref.
\cite{kn:girvin1}-\cite{kn:jain1}) is to first find a Mean Field Solution
to the above equations. The properties of the Quantum system can then be
studied by expanding about this mean field.
We thus begin by solving the equations of motion 
(\ref{eq:motion1}-\ref{eq:motion3})
as if they were classical equations. When 
\equation
	\frac{\mu}{\lambda}~=~\frac{eB}{2\pi n}
	\label{eq:concon}
\endequation
this set of equations has the constant solution 
$\vert\psi\vert^2=\frac{\mu}{\lambda}$, $e_j=0$, and $gb=eB$. If either
$\mu$ or $B$ is changed so that Eq. (\ref{eq:concon}) is not satisfied
it would seem difficult to find solutions which simultaneously
minimize the potential and produce no net field. We now analyze the
equations in this case in a more systematic manner.

We begin with several important observations about these equations 
which are generalizations of  theorems proven in Ref \cite{kn:jawe}
in a somewhat different context. The first observation is that 
unless $\mu/\lambda=eB/(2\pi n)$ (which corresponds to the
special filling fraction)
there are $no$ nonzero
covariantly constant solutions (solutions with $D_\mu\psi=0$)
to this system of equations.  The proof is as follows:
first write $\psi=\xi {\rm e}^{i\Omega}$. 
Now $D_\mu\psi=0$ implies that $\partial_\mu\xi=0$ and that
$\partial_\mu\Omega=eA_\mu-ga_\mu$ so that the combination $eA-ga$ is a 
pure
gauge. It follows that $eB=gb$. Now Eqs. (\ref{eq:motion1}) and 
(\ref{eq:motion2}) with $D_\mu\psi=0$ imply that either $\psi=0$
or that $\mu/\lambda=eb/(2\pi n)$
so that $\mu/\lambda=eB/(2\pi n)$. This completes the proof that nonzero 
covariantly constant solutions only exist when $\mu/\lambda=eB/(2\pi n)$.

In addition to the lack of solutions with $D_\mu\psi=0$ there are also no nonzero
covariantly static solutions (i.e. solutions with $D_0\psi=0$) unless
$\mu/\lambda=eB/(2\pi n)$. The proof again begins by writing 
$\psi=\xi {\rm e}^{i\Omega}$.
$D_0\psi=0$ then implies that $\partial_0\xi=0$
and that $\partial_0\Omega=eA_0-ga_0=-ga_0$. Using this and the third equation
of motion Eq. (\ref{eq:motion3}) as well as the fact that 
$\partial_0 A_i=0$ we find that
\equation
	\partial_0 \left(\partial_j\Omega+ga_j-eA_j\right)
	\propto \epsilon_{jk}\left(\partial_k\Omega+ga_k-eA_k\right)\xi^2
	\label{eq:prop}
\endequation
Since $\xi$ is independent of time we can identify three possibilities.
The first option is  $\xi=0$ in which case $\psi=0$. A second possibility
is that $\partial_j\Omega+ga_j-eA_j=0$ in which case $gb=eB$. In this case
Eq. (\ref{eq:motion2}) (the Chern--Simons constraint) implies that
$\vert\psi\vert^2=eB/2\pi n$ which is the condition for the special 
filling
fraction. This further implies that $\xi$ is spatially constant which,
from Eq. (\ref{eq:motion1}) forces $\vert\psi\vert^2=\mu/\lambda$.
The third possibility for satisfying Eq. (\ref{eq:prop}), which is 
necessary
when we have a nonzero $\psi$ with $\mu/\lambda\ne eB/(2\pi n)$ is that
$\partial_j\Omega+ga_j-eA_j$ is a nonzero oscillating (sinusoidal) 
function of time. It follows that $gb-eB$ and thus $gb$ is an oscillating 
function of time which then implies that $\xi$ oscillates with time 
which contradicts the equation $\partial_0\xi=0$. It follows that
except at special filling fractions there is no nonzero solution
which is covariantly static.

Despite the fact that no covariantly static solution exists it 
$is$  possible to find nonzero time independent solutions 
(with $\partial_0\psi=0$) but with $a_0\ne 0$.  As is shown (in a somewhat
different context) in Ref. (\cite{kn:jawe}) this can be done by extremizing
the energy functional
\equation
	H=\int d^2x ~{\cal H} =\int d^2x \left[
	\frac{1}{2m}\vert D_i\psi\vert^2 - \mu \vert\psi\vert^2
	+ \frac{\lambda}{2}\vert\psi\vert^4\right]
	\label{eq:hamton}
\endequation
with $D_i=\partial_i-ieA_i+iga_i$ but
subject to to the constraint Eq. (\ref{eq:motion2}) that
\equation
	\nabla \times a = \frac{2\pi n}{g} \vert\psi\vert^2
	\label{eq:hamconst}
\endequation
In other words when Eq. (\ref{eq:hamton}) is varied with respect to $\psi$
and $a_i$ subject to the constraint (\ref{eq:hamconst}) one obtains
precisely the equations of motion (\ref{eq:motion1}--\ref{eq:motion3})
with $\partial_0\psi=0$ but with $a_0\ne 0$. The term proportional to $a_0$
appears when $a_i$ is varied with respect to $\psi$ as required by
the constraint Eq. (\ref{eq:hamconst}). $a_0$ will be a solution to 
Eq. (\ref{eq:motion3}) with $\partial_0 a_i=0$ (which can always be arranged
since $\psi$ is time independent).

Our job is now to find a configuration of $\psi$ and $a_i$ which 
minimizes the energy (\ref{eq:hamton}) subject to the constraint 
(\ref{eq:hamconst}) for various values of the parameters $\mu$ and $B$.
The simplest case which has already been discussed is the case in which 
$\mu/\lambda=eB/(2\pi n)$. In this case the configuration with 
$\rho=\mu/\lambda$ both minimizes the potential and forces
$eb=gB$ which admits a configuration with $D_i\psi=0$.  
In this situation the density $\rho$ satisfies
$\nu=2\pi\rho/eB=1/n$ which is the special filling fraction.
The case of interest to us
however occurs when either $B$ or $\mu$ is modified so that 
\equation
	\frac{eB}{2\pi n} = f\frac{\mu}{\lambda} \ne \frac{\mu}{\lambda}
	\label{eq:nomagic}
\endequation
In this case there is a problem. If we force the
potential to be at its minimum $\vert\psi\vert^2=\mu/\lambda$ then
$eB\ne gb$ and the $\vert D_i\psi\vert^2$ term in the Hamiltonian 
(\ref{eq:hamton})
leads to an infinite energy (the Meissner effect). If, on the other hand,
we force $eB=gb$ (which must then be spatially constant) then 
$\vert\psi\vert^2=f\mu/\lambda$ which is not at the minimum of the 
potential.

The solution to the above problem is as follows: Suppose we first consider
the configuration described above in which $\vert\psi\vert^2=f\mu/\lambda$
is
$not$ at the minimum of the potential. It may be surprising, but the fact
is that this configuration $is$ an extremum of the Hamiltonian 
(\ref{eq:hamton})
subject to the constraint (\ref{eq:hamconst}). To see this notice that
this configuration $does$ satisfy the equations of motion 
(\ref{eq:motion1}--\ref{eq:motion3}) but with 
\equation
	ga_0~=~ \frac{\mu^2}{\lambda}f(1-f)
\endequation
which is spatially constant. We see from Eq. (\ref{eq:motion1})
that the constant part of $a_0$ behaves in the same manner as a shift
in the chemical potential thus allowing the minimum of the potential
to be consistent with the cancellation of $eB$ and $gb$. 

It is also clear
intuitively why this configuration is an extremum of the Hamiltonian 
(\ref{eq:hamton}). The main candidate for a fluctuation which might
lower the energy is one in which $\psi$ is changed locally by an
infinitesimal amount in the direction of the minimum of the potential.
This will cause an infinitesimal amount of flux ($\int(eB-gb)$) to
thread through the system. If this flux is truly infinitesimal
it will be less than one flux quantum and thus its gauge potential will
lead to an infinite contribution to the gradient term in the Hamiltonian.
(It is of course possible to consider fluctuations which have zero
total flux but there is no reason to believe that these would lower the
energy. The fact that they leave the energy unchanged to lowest order
is evidenced by the previous argument that this configuration solves
the equations of motion.) 

The configurations in which the flux is not infinitesimal but rather
consists of an integral number of flux quanta $is$ a finite energy
excitation of this system. If, for example, we imagine modifying
the chemical potential away from the special value we might hope that
these vortices would lower the energy of the system and thus describe
the Fractional Quantum Hall system $away$ from special filling as
a collection of these vortices. Unfortunately this is not the case.
We shall see shortly that for
small deviations of $\mu$ from its ``special'' value these vortices
always $increase$ the energy. Let us first discuss why this is the
case and then explain its consequences.

The main point is that the configurations which are extrema of the 
Hamiltonian (\ref{eq:hamton}) subject to the constraint (\ref{eq:hamconst})
are completely independent of the value of the chemical potential $\mu$!
We have, in fact, proven this already. Configurations  are
extrema of (\ref{eq:hamton}) subject to (\ref{eq:hamconst}) if and
only if they satisfy the equations of motion 
(\ref{eq:motion1}--\ref{eq:motion3}). Thus if a particular $\psi$ is such
an extremum (it may be a constant or, more generally, a multivortex
configuration)
with a given value $\mu_1$ of $\mu$ then it will satisfy
the equations of motion for some function $a_0^{(1)}$. The same $\psi$
will satisfy the equations of motion for any other value $\mu_2$
of $\mu$ but this time with a new $a_0^{(2)}=a_0^{(1)}+\mu_2-\mu_1$.
It is thus also an extremum of the Hamiltonian with this new value
$\mu_2$ of $\mu$. 

Even though the extremal configurations for differing values of $\mu$
are the same, the energetics may differ for different values of $\mu$.
Notice, for example, that for $\mu$ such that $\mu/\lambda=eB/(2\pi n)$
(let us call this value of the chemical potential $\mu_0$),
the configuration $\vert\psi\vert^2=\mu_0/\lambda$ clearly has the 
lowest possible energy. A single vortex configuration $\psi_v(x)$ which 
solves the equations of motion (see \cite{kn:vortsl}) will have a 
larger energy than the ground state. Let $\epsilon_v$ be the excess energy
of the vortex with respect to the ground state of the 
Hamiltonian with $\mu=\mu_0$.
Now consider an alternate Hamiltonian with $\mu=\mu_1\ne\mu_0$.
The same configuration $\psi_v(x)$ will still be an extremum of this
Hamiltonian but its energy $\epsilon_v^{(1)}$ (which is the
{\it difference} in energy between the vortex configuration and
the configuration with $\vert\psi\vert^2=\mu_0/\lambda$) will differ:
\equation
	\epsilon_v^{(1)}=\epsilon_v-(\mu_1-\mu_0)
	\int d^2x 
	\left(\vert\psi_v(x)\vert^2-\frac{\mu_0}{\lambda}\right)
	=\epsilon_v-(\mu_1-\mu_0)N_v
	\label{eq:newps}
\endequation
where $N_v=\pm 1/n$ is the ``particle number'' of the vortex.
Note that when $\mu$ is decreased it is preferable to form a vortex
($N_v<0$) whereas if $\mu$ is increased an antivortex is preferred.

Equation (\ref{eq:newps}) has the following consequences. 
{\bf For small values} of $\delta\mu=\mu_0-\mu_1$ the vortex 
configuration {\it increases} the energy of the system at $\mu=\mu_1$.
We thus expect that the constant configuration with 
$\vert\psi\vert^2=\mu_0/\lambda$ will be the configuration of lowest
energy despite the fact that the potential energy is not at its 
minimum. This configuration has a density $\rho=\mu_0/\lambda$ so
that the system remains at the special filling fraction even after 
$\mu$ has been shifted from $\mu_0$ to $\mu_1$. 
(We emphasize again that this occurs for small shifts $\mu_0-\mu_1$.)
It follows that
in the Mean Field approximation
\equation
	\frac{d\rho}{d\mu}_{\vert_B}~=~0
\endequation
for a range of $\mu$ near $\mu=\mu_0$. This equation is familiar from
the integer Quantum Hall effect and is  due to the presence
of a gap in the spectrum. (See Refs.
 \cite{kn:gros} for a discussion of 
this with respect to the FQHE). It implies that a finite change in 
the chemical potential is required before the density can be modified.
Returning to Eq. (\ref{eq:newps}) we see that when 
$\delta\mu=\vert\mu_0-\mu_1 \vert\ge \epsilon_v/N_v$ a single vortex has 
$lower$ energy than the constant configuration
$\vert\psi\vert^2=\mu_0/\lambda$. A weakly interacting gas of such
vortices will have an even lower energy. We thus expect that
near this value ($\epsilon_v/N_v$) of $\delta\mu$ the lowest energy
configuration of the Hamiltonian (\ref{eq:hamton}) subject to the 
constraint (\ref{eq:hamconst}) will be a collection of vortices 
(which will likely form a lattice in the Mean Field approximation).

The situation is similar when  the magnetic field is modified
instead of $\mu$. If we begin at the special filling fraction
with $eB=2\pi n\mu_0/\lambda$ and change $B$ at fixed $\mu$ by
a small amount,
the lowest energy configuration of the Hamiltonian will occur
at a $new$ density for which $2\pi\rho/eB$ is still equal to $1/n$
but which will now $not$ be at the minimum of the potential.
As the magnetic field is increased (or decreased) further
(again at fixed $\mu=\mu_0$) the cost in energy of a single
vortex (or antivortex) becomes progressively smaller until 
at some critical value of the field it becomes negative. At that
point the lowest energy mean field configuration
is no longer a constant but rather a lattice of vortices
in which case $2\pi\rho/eB$ is no longer equal to $1/n$. 
(If the magnetic field is increased then the vortices will
``condense'' whereas if it is decreased the antivortices
will ``condense''.)
Thus if $2\pi\rho/eB$ is plotted either as a function of $\mu$
or as a function of $B$ there is a plateau surrounding the
value $\mu_0$ for which this ratio is constant and equal to
$1/n$. If, on the other hand, the $B$ is varied at fixed $density$
then we move off the plateau and the lowest energy mean
field configuration consists of a vortex lattice.

In the remainder of this paper we shall look more closely at the
vortices of this model which, as we have discussed, will be
solutions both for $\mu=\mu_0$ and for values of $\mu$ differing
from $\mu_0$. We shall present some numerical solutions for these 
vortices which will allow us to estimate the value of
$\mu$ at which a lattice of vortices starts to form. The shape
of the vortices will also lead to an estimate of the density
at which the collection of vortices becomes non--dilute.
The method we have chosen for finding vortices is 
by considering the
Hamiltonian and constraint given by Eqs. (\ref{eq:hamton}--\ref{eq:hamconst})
at a value of $\mu = \mu_0 = eB\lambda/(2\pi n)$. We then look for
radially symmetric configurations (which necessarily carry an
integer number of flux quanta of $eB-gb$) which minimize the
Hamiltonian.

Anticipating the fact that our  solution will be a vortex with
an integer number of flux quanta we organize a radial ansatz  
as follows: First write
\equation
	\psi(r,\theta)=\xi(r){\rm e}^{-ik\theta}
	\label{defpsi}
\endequation
with $k$ an integer. The Hamiltonian (\ref{eq:hamton}) can
now be written as:
\equation
	H=\int_0^\infty 2\pi rdr\left[
	\frac{1}{2m}\left\{ \left(\frac{d\xi(r)}{dr}\right)^2
	+\left(\frac{k}{r}+\frac{eBr}{2}
	-\frac{g}{r}\int_0^r \hat r b(\hat r)d\hat r\right)^2\xi^2(r)\right\}
	-\mu\xi^2(r)+\frac{\lambda}{2}\xi^4(r)\right]
	\label{radham}
\endequation
with the constraint
\equation
	\xi^2(r)=\frac{gb(r)}{2\pi n}
	\label{radconst}
\endequation
(Note that if we were considering a value of $\mu$ not equal to $\mu_0$
we would still use the above equation but  with $\mu=\mu_0/f$
so that the Mean Field Solution would be $\xi^2=f\mu/\lambda$.)
We now define the function $h(r)$ via the formula
\equation
	gb(r)-eB=\frac{h^\prime(r)}{r}
	\label{ansz}
\endequation
where $h^\prime(r)=dh/dr$. $h(0)$ can be chosen equal to $0$
without loss of generality. The second term in Eq. (\ref{radham}) is then
proportional to:
\equation
	\left(\frac{k}{r}+\frac{eBr}{2}
	-\frac{g}{r}\int_0^r r^\prime b(r^\prime)dr^\prime\right)^2\xi^2(r)
	=\left(\frac{k}{r}-\frac{h(r)}{r}\right)^2\xi^2(r)
	\label{radhamsim}
\endequation
In order for the integral to be finite at large $r$ we require 
$h(\infty)=k$ (or, more precisely, we require it to be an integer
and we choose $k$ in Eq. (\ref{defpsi}) to be that integer).
(We only consider the cases $k=\pm 1$ in this paper, since those
configurations have the lowest energy.)
Furthermore, as is standard for all vortices, $\xi^2(r)$ must vanish
at the origin in order for the energy to be finite. The Chern--Simons
condition Eq. (\ref{radconst}) then implies
\equation
	\xi^2(r)=\frac{gb}{2\pi n}=\frac{eB+h^\prime(r)/r}{2\pi n}  ~\ge 0 ~~ 
	{\rm and} ~
	\rightarrow 0 ~~{\rm as}~~r\rightarrow 0
	\label{xidef}
\endequation
This is the most difficult condition to implement in a numerical scheme
in which the function $h(r)$ is varied to minimize the energy.

With the above definitions the Hamiltonian is given by
\equation
	H=2\pi\int_0^\infty rdr \left[ 
	\frac{1}{2m}\left(\frac{d\xi(r)}{dr}\right)^2
	+\frac{1}{2m}
	\left(\frac{k}{r}-\frac{h(r)}{r}\right)^2\xi^2(r)
	-\mu\xi^2(r)+\frac{\lambda}{2}\xi^4(r)\right]
	\label{finham}
\endequation
with $h(r)$ chosen so that $h(0)=0$, $h(\infty)=k$ and $\xi^2(r)$, defined
by Eq. (\ref{xidef}), is $\ge 0$. Notice that these conditions guarantee
that the total flux of the vortex 
\equation
	\int d^2x \left(gb-eB\right)=2\pi k
	\label{fluxx}
\endequation
The final step is to subtract, from the energy of the vortex solution
of Eq. (\ref{finham}), the energy of the Mean Field solution
$\xi^2=f\mu/\lambda$. This results in a vortex energy given by:
\begin{eqnarray}
	H_{\rm v}& =& 2\pi\int_0^\infty rdr \left[ 
	               \frac{1}{2m}\left(\frac{d\xi(r)}{dr}\right)^2
	              +\frac{1}{2m}
                	\left(\frac{k}{r}
                       -\frac{h(r)}{r}\right)^2\xi^2(r)\right. \nonumber \\
	  &   & \left.
	-\mu(1-f)\left(\xi^2(r)-f\frac{\mu}{\lambda}\right)
	+\frac{\lambda}{2}\left(\xi^2(r)-f\frac{\mu}{\lambda}
	\right)^2\right]
	\label{finfinham}
\end{eqnarray}
The procedure at this stage is to 
search, numerically, through the space of such functions $h(r)$
until the Hamiltonian is minimized.

One point which is clear is that the form of the vortex solution
(for which $\int(eB-gb)>0$ is quite different from that of the 
antivortex solution. The reason is that $\xi(r)$ and thus $b(r)$ must
vanish at the origin. As a consequence the density $\psi^2(r)\propto
b(r)$ for
the vortex solution  can be a monotonic function of $r$ which increases
from $0$ at the origin and reached $eB$ at infinity. The antivortex 
solution must however be zero at the origin then increase to
a value greater than $eB$ (so that $\int(eB-gb)<0$) and then
decrease again to attain its asymptotic value $eB$ as $r\rightarrow\infty$.
We shall of course see this behavior clearly in the numerical
solutions shown in Figures 1-4 below.

For the numerical work we chose 
some representative values for the parameters of 
the model \cite{kn:expt}:
\begin{eqnarray}
    \mu_0 & = & .010 \, {\rm eV} \nonumber \\
    \mu_0/\lambda & = & 10^{3} \,\mu{\rm m^{-2}} \nonumber \\
    m & = & .08 \, m_{e} 
\end{eqnarray}
(In the units $\hbar=c=1$ this translates to $\lambda=.00025\,{\rm eV}^{-1}$,
$\rho=\mu_0/\lambda=40\,{\rm eV}^2$,
$eB=250n\,{\rm eV}^2$ and $m=41000\,{\rm eV}$.) 
Note that $\mu_0/\lambda$ is the density of carriers. (Although
the above values were chosen to be representative of  experiments
which exhibit the FQHE it is 
difficult to call them ``realistic'' since the CSLG
model has been greatly simplified by replacing the Coulomb
interaction of the charge carriers with a short range interaction.)

\begin{figure}[ht]
\epsfysize=6.0in
\epsfbox{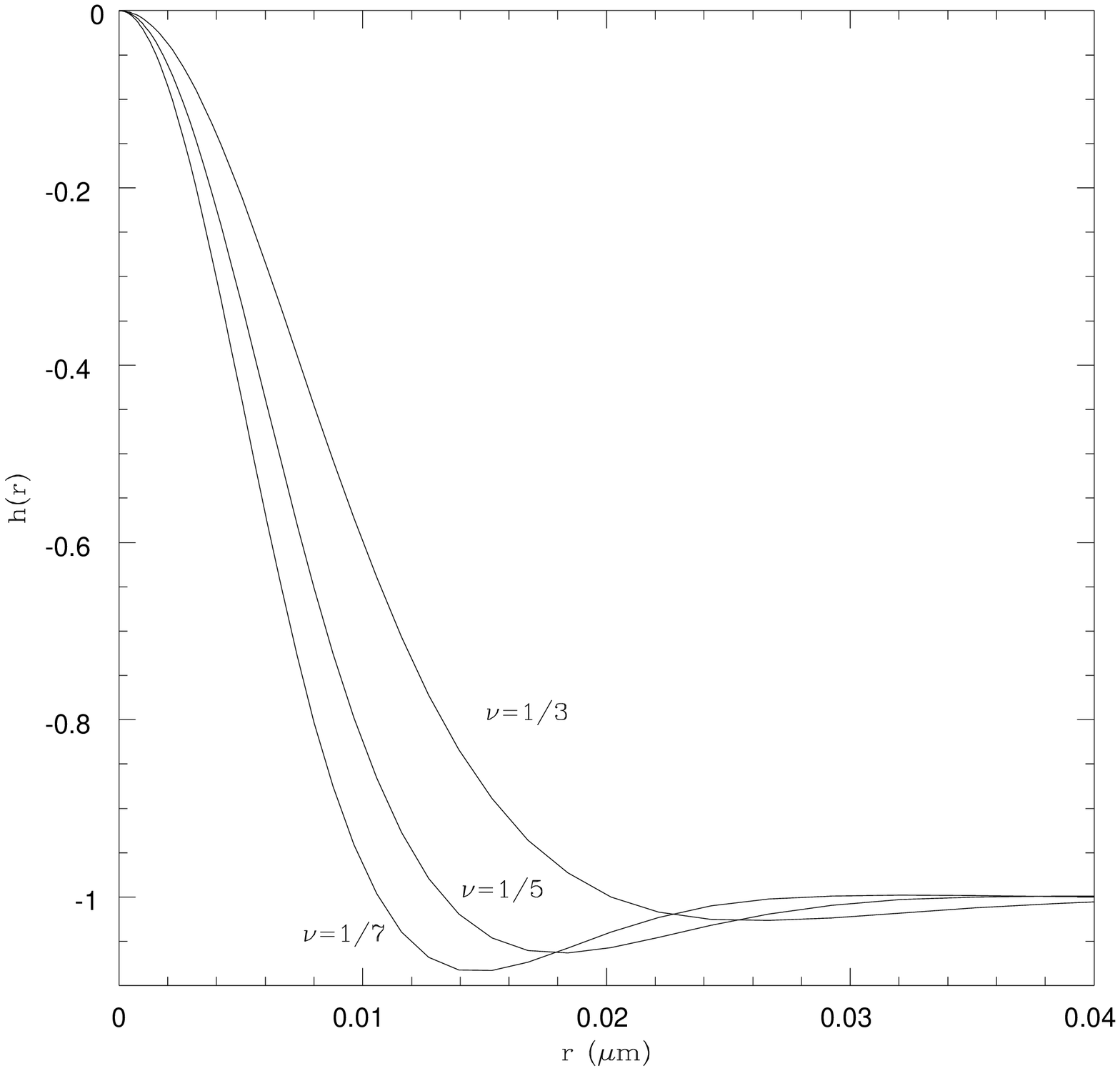}
\caption{The function $h(r)$ corresponding to a vortex 
for $\nu=1/3,1/5$ and $1/7$. \label{fi:hminus}}
\end{figure}

\begin{figure}[ht]
\epsfysize=6.0in
\epsfbox{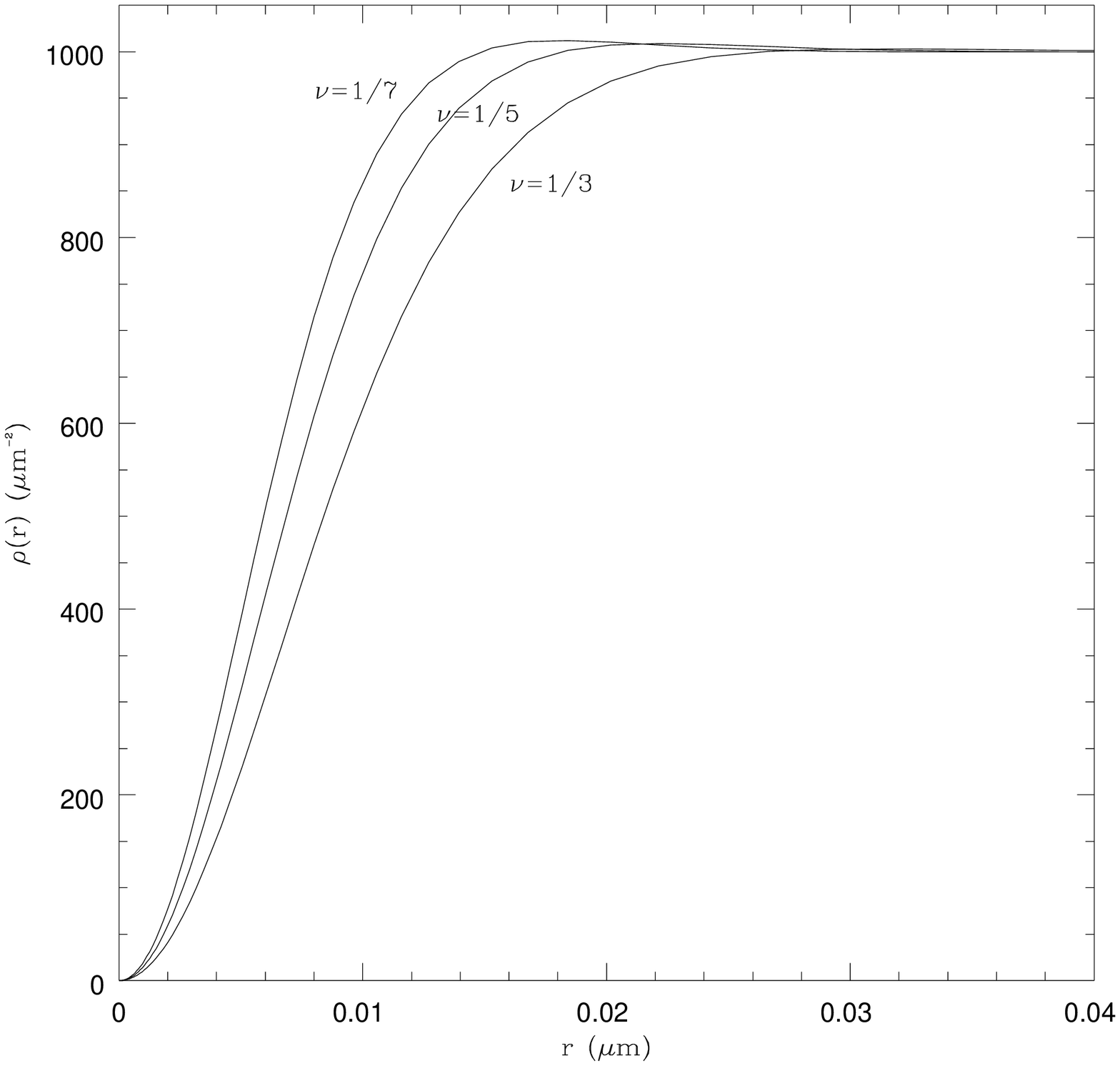}
\caption{Density profile of the vortex configuration for 
$\nu = 1/3,1/5$ and $1/7$. \label{fi:dvortex}}
\end{figure}

\begin{figure}[ht]
\epsfysize=6.0in
\epsfbox{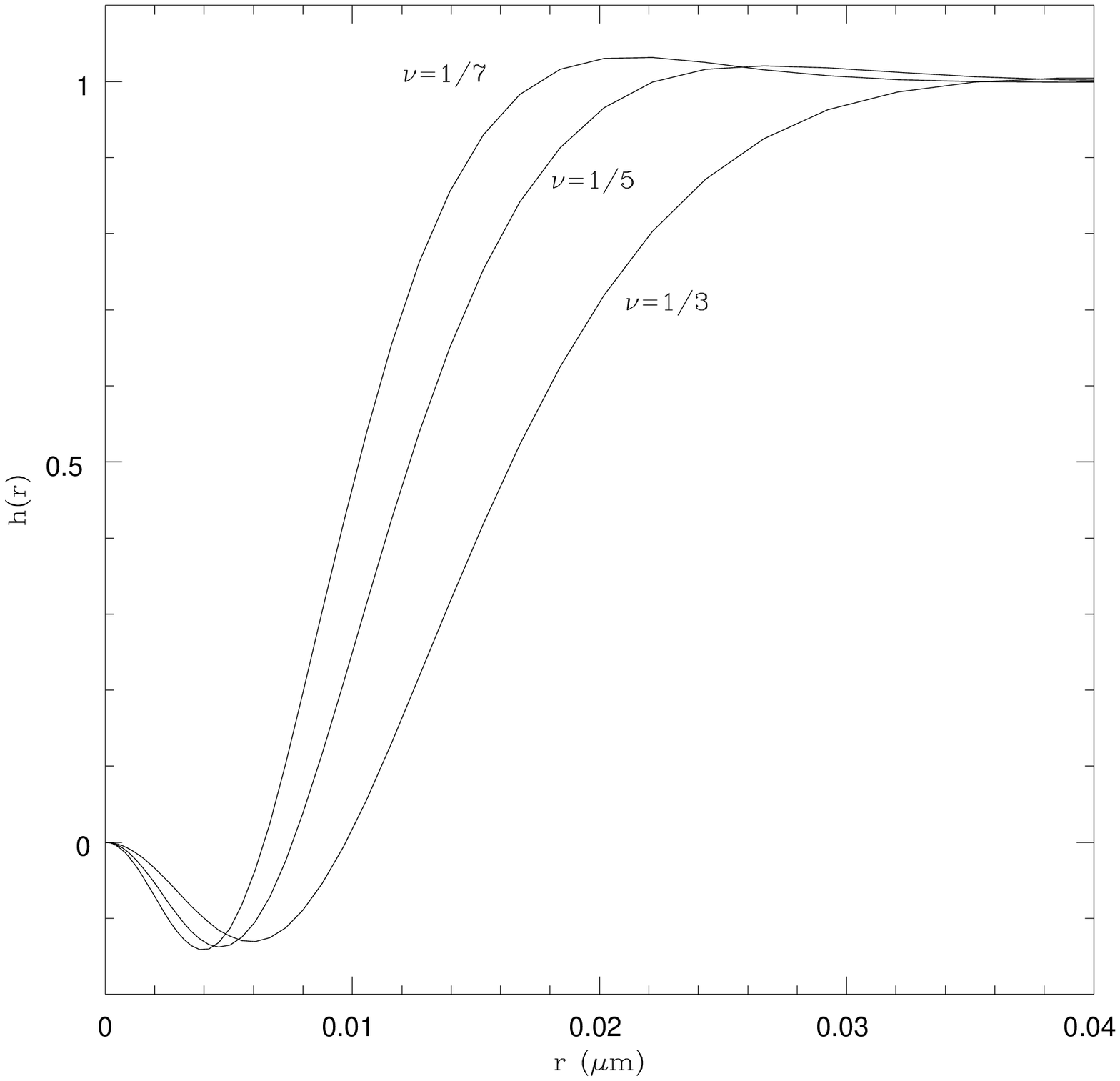}
\caption{The function $h(r)$ corresponding to an antivortex 
for $\nu=1/3,1/5$ and $1/7$. \label{fi:hplus}}
\end{figure}

\begin{figure}[ht]
\epsfysize=6.0in
\epsfbox{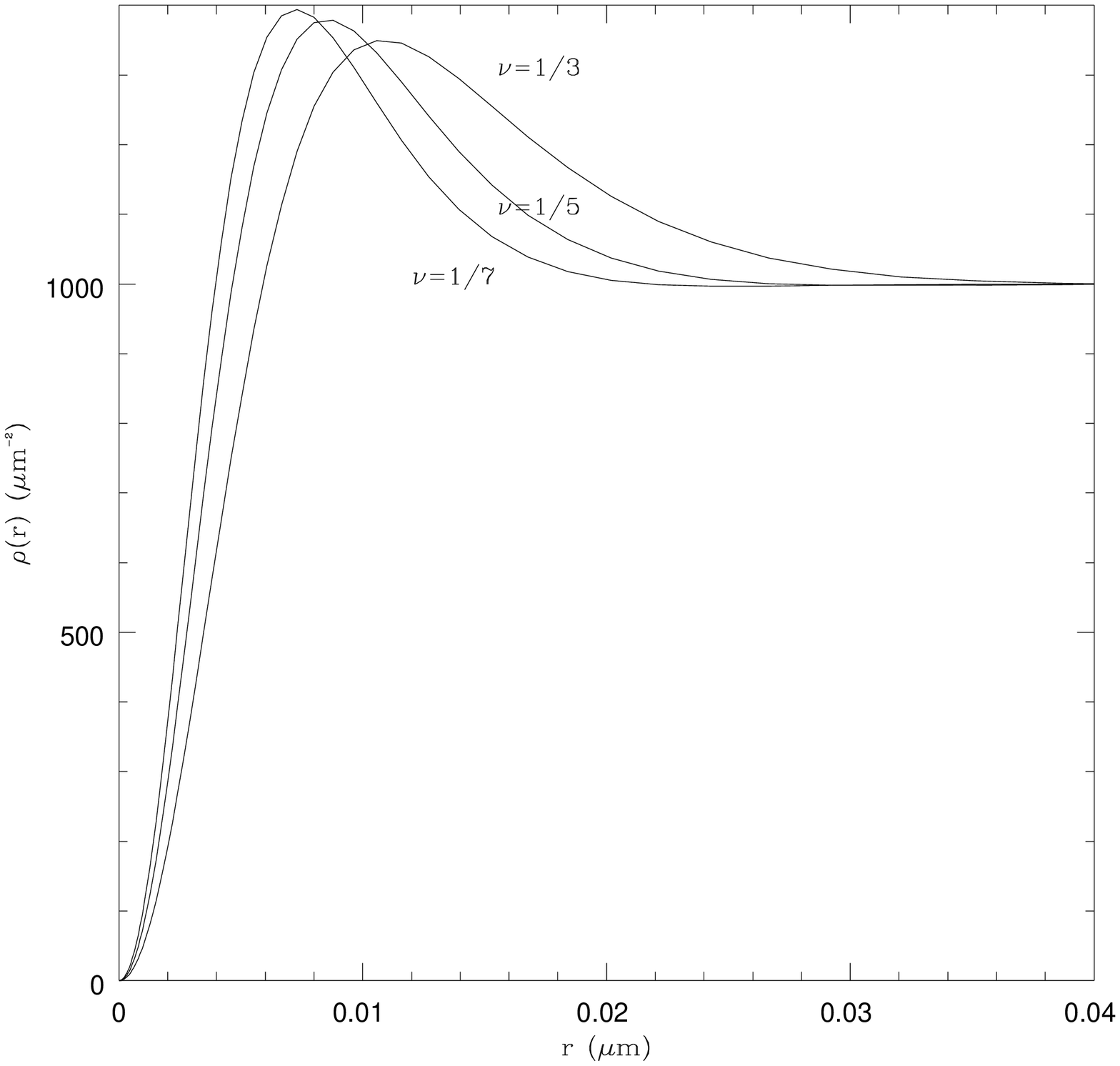}
\caption{Density profile of the antivortex configuration for 
$\nu = 1/3,1/5$ and $1/7$. \label{fi:davortex}}
\end{figure}

In Fig. \ref{fi:hminus} and Fig. \ref{fi:dvortex} 
we present the numerical solution for the
{\bf vortex} configurations (in which $gb$ is lowered relative
to $eB$ and $h(r)\rightarrow -1$ as $r\rightarrow \infty$). 
We plot the functions $h(r)$ and the density
$\rho(r)$ for $\nu=1/3,1/5$ and $1/7$ respectively. When changing
$\nu$, the density $\rho$ remains fixed as the magnetic field
$B$ is varied. Figures \ref{fi:hplus} and \ref{fi:davortex} 
contain plots of $h(r)$ and $\rho(r)$
for the same values of $\nu$ but now for
the {\bf antivortex} configurations.
In Table~\ref{ta:energy} we present the energies and some measure $r_0$ of
the size of each vortex and antivortex. We have arbitrarily
chosen the size of the vortex as the value of $r$ at which
the energy density has reached $99\%$ of its total value.

We are now ready to describe quantitatively (within this model)
what happens when the chemical potential is varied from $\mu_0$.
As discussed in great detail in this paper there is no change
in the density unless $\mu-\mu_0$ is approximately equal to 
the energy of a vortex times $n$ (i.e. the energy per particle 
of the vortex). We can now see, quantitatively, how this works from
Eq. (\ref{finfinham}). If $f\ne 1$ so that $\mu=\mu_0/f\ne\mu_0$
then the energy of the vortex is simply
\equation
	\epsilon_v(\mu)=\epsilon_v(\mu_0)-\left(\mu-\mu_0\right)
	\times\left(\mp\frac{1}{n}\right)
\endequation
where the minus sign is for a vortex and the plus sign for an
antivortex. Thus for $\mu<\mu_0$ the vortex configuration has lower
energy than the antivortex configuration. We naively expect that
when $\mu_0-\mu=n\epsilon_v(\mu_0)$ (or near this point) the Mean
Field ground state should be a condensate (possibly a lattice) of 
vortices. Conversely when $\mu>\mu_0$ the antivortex
has lower energy and when $\mu-\mu_0=n\epsilon_v(\mu_0)$ we naively
expect a condensate of antivortices. Unfortunately, for our choice
of parameters, $\epsilon_v$ is quite large. Thus
the value  $\mu_{cr}$ or $\mu$ at which this condensate occurs 
in the above naive calculation and which is  shown in Table 
\ref{ta:energy} differs from $\mu_0$ by an unreasonably large
amount. This leads, in particular, to a negative value of
$\mu_cr$ for the $n=5$ and $n=7$ vortex configurations.
In fact, as $\mu$ is varied from $\mu_0$ towards $\mu_{cr}$,
another critical value $\hat\mu$ of $\mu$ is reached at which
$2\pi\hat\mu/\lambda eB=1/(n+2)$ well before $\mu_{cr}$ is
reached. At $\hat\mu$ the system is better described
by a Chern--Simons theory with the new value $\hat\nu=1/(n+2)$
of the filling fraction.

In light of the above remarks we should try to understand whether
in fact one does form a vortex condensate in our model at our
chosen values of the parameters. Certainly the vortices must
be strongly interacting (or overlapping) well before $\hat\mu$
or $\mu_{cr}$ is reached. This can be better understood by first
supposing 
that such a condensate {\bf} is formed as $\mu$ is lowered.
The approximate density $\rho_1$ at which a description of 
this condensate in terms of the single vortex solutions 
presented above fails
depends most prominently on the size of
a vortex. In Table \ref{ta:two} we show the approximate density
of vortices $\rho_{v1}$
and chemical potential $\mu_1$ at which the vortices begin to
``touch''. For a hexagonal lattice of vortices this will occur
when $\rho_{v1}=1/(2\sqrt{3}d^2)$, ie, where d (the distance between
vortices) equals the size of the vortices.
Table \ref{ta:two} also shows the corresponding values of 
the filling fraction. Notice that in most cases the vortex condensate
becomes dense well the before the ``next'' value of $n$ (i.e. 
well before $2\pi\rho/eB=1/(n\pm 2)$). 
We conclude from this that an approximation in terms of
a dilute gas of vortices breaks down well before 
$\mu=\mu_{cr}$. It is thus likely that even for our chosen
values of the parameters a condensate of vortices will form
in the Mean Field description.
The formation of this vortex condensate and the resultant pinning
of the vortices is what gives rise to the 
hall plateaus by allowing the system to continue to behave as if it were
in a state $\nu=1/n$ even after the magnetic field or the chemical potential 
has been changed to move it away from that value. 
This is because the (anti)vortices
accommodate the localized excess (deficit) of charge.  

\vskip .4in

\centerline{\bf Summary}
\medskip

In this paper we have studied the Mean Field behavior of the
CSLG description of the FQHE when the filling fraction 
deviates form the ``special'' filling fraction for which
$\nu=1/n$ with $n$ and odd integer. We have shown how the
Field Theoretic description of this model at a fixed
chemical potential $\mu$ and magnetic field $B$ can be
studied for a range of $\mu$ surrounding the  value
$\mu_0$ corresponding to the special filling fraction.
For small values of $\vert\mu-\mu_0\vert$ (and at zero
temperature) the density is independent of $\mu$. This is 
reminiscent of what occurs for the integer Quantum Hall Effect.
As $\mu$ is decreased beyond some $\mu_{cr}^-$ we show how
the homogeneous Mean Field configuration is unstable to the
formation of a condensate of vortices. If $\mu$ is increased
above some $\mu_{cr}^+$ the instability is to the formation
of a condensate of antivortices.  We have presented a
numerical example of these vortex and antivortex configurations
and we estimated the densities and filling fractions 
at which the description in terms of a noninteracting system
of vortices breaks down.

\vfill
\eject

\begin{table}[ht]
\begin{center}
\begin{tabular}{|c|c|c|c|c|}  \hline
vortex/antivortex    & $n$   &size ($\mu m$) & energy ($\times 10^{-2}eV$)
& $\mu_{cr}$ ($\times 10^{-2}eV$)\\ \hline\hline
      		     &  3    &0.027       &  0.25  & 0.25 \\ \cline{2-5}
vortex               &  5    &0.018       &  0.21  & -0.05 \\ \cline{2-5}
                     &  7    &0.015       &  0.20  & -0.4 \\ \hline
                     &  3    &0.035       &  0.83  & 3.5 \\ \cline{2-5}
antivortex           &  5    &0.025       &  0.80  & 5.0 \\ \cline{2-5}
                     &  7    &0.020       &  0.79  & 6.5 \\ \hline
\end{tabular}
\end{center}
\caption{ Energy and size ($r_0$) of vortices and antivortices
for $n=3,5$ and $7$. $\mu_{cr}$ is a naive estimate of 
the value of the chemical
potential at which
a condensate of these configurations is expected to form
($\mu_0=10^{-2}ev$). $\mu_cr$ is more carefully described in the
text.
\label{ta:energy}}
\end{table} 

\vskip .4in

\begin{table}[ht]
\begin{center}
\begin{tabular}{|c|c|c|c|c|}  \hline
vortex/antivortex    & $n$   & $\rho_{v1}$ ($\mu m^{-2}$)
& $\nu_1^{-1}$\\ \hline\hline
      		     &  3       &  400  &2.6\\ \cline{2-5}
vortex               &  5       &  890  & 4.1  \\ \cline{2-5}
                     &  7       &  1280  & 5.7 \\ \hline
                     &  3       &  240  & 3.2\\ \cline{2-5}
antivortex           &  5       &  460  & 5.5 \\ \cline{2-5}
                     &  7       &  720  & 7.7 \\ \hline
\end{tabular}
\end{center}
\caption{ The density $\rho_{v1}$
and the corresponding
filling fraction $\nu_1$ at which the
condensate of vortices is expected to become dense. (See text for a precise
definition.)
\label{ta:two}}
\end{table}

\vskip .4in
 
\end{document}